\documentclass[11pt]{article}

\usepackage[final]{acl}
\usepackage{makecell}
\usepackage{times}
\usepackage{latexsym}
\usepackage{tabularx}

\usepackage[T1]{fontenc}

\usepackage[utf8]{inputenc}

\usepackage{microtype}

\usepackage{inconsolata}

\usepackage{amsmath,amssymb}
\usepackage{booktabs}
\usepackage{tabularx}
\usepackage{enumitem}
\usepackage{tikz-cd}
\usepackage{multicol}
\usepackage{framed}
\usepackage{todonotes}

\title{MLOps-Bench: Benchmarking AI Agents on Production ML Engineering in Real Repositories}
\author{
Quoc Co Tran, Hitesh Laxmichand Patel, Diwakar Mahajan \\
\textbf{Kshitij Bakliwal, Avi Sil, Katrin Kirchhoff}\\
Oracle \\
\texttt{\{co.tran,hitesh.laxmichand.patel,diwakar.mahajan,}\\
\texttt{kshitij.bakliwal,avi.sil,katrin.kirchhoff\}@oracle.com}
}
\begin{document}
\maketitle

\begin{abstract}
AI agents are increasingly expected to modify and maintain machine-learning systems in production, yet existing benchmarks primarily evaluate either competition-style ML experimentation or generic repository issue resolution. We introduce \textsc{MLOps-Bench}, a repository-grounded benchmark for production ML engineering. Unlike prior ML-agent benchmarks, \textsc{MLOps-Bench} evaluates executable lifecycle changes inside real ML repositories, spanning data pipelines, feature engineering, model training, model serving, monitoring, CI/CD and governance, and end-to-end integration. Each task is grounded in repository evidence and includes a natural-language requirement, oracle patch, plausible buggy baseline, fail-to-pass tests, pass-to-pass tests, and provenance metadata. The benchmark contains 212 validated tasks from 31 public ML repositories, with 636 fail-to-pass tests and 856 pass-to-pass tests. We evaluate six agents and find that the best system solves 31.2\% of tasks while passing 68.7\% of tests, showing that current agents often make partial progress but still fail production-relevant behavioral requirements. \textsc{MLOps-Bench} \footnote{The evaluation code and benchmark tasks are released at https://github.com/oracle-samples/mlops-bench.}  shifts agent evaluation beyond model building and issue fixing toward repository-native MLOps work where correctness depends on lifecycle reasoning, behavioral robustness, and preservation of existing system behavior.
\end{abstract}

\begin{figure*}[!th]
\centering
\includegraphics[width=0.95\linewidth]{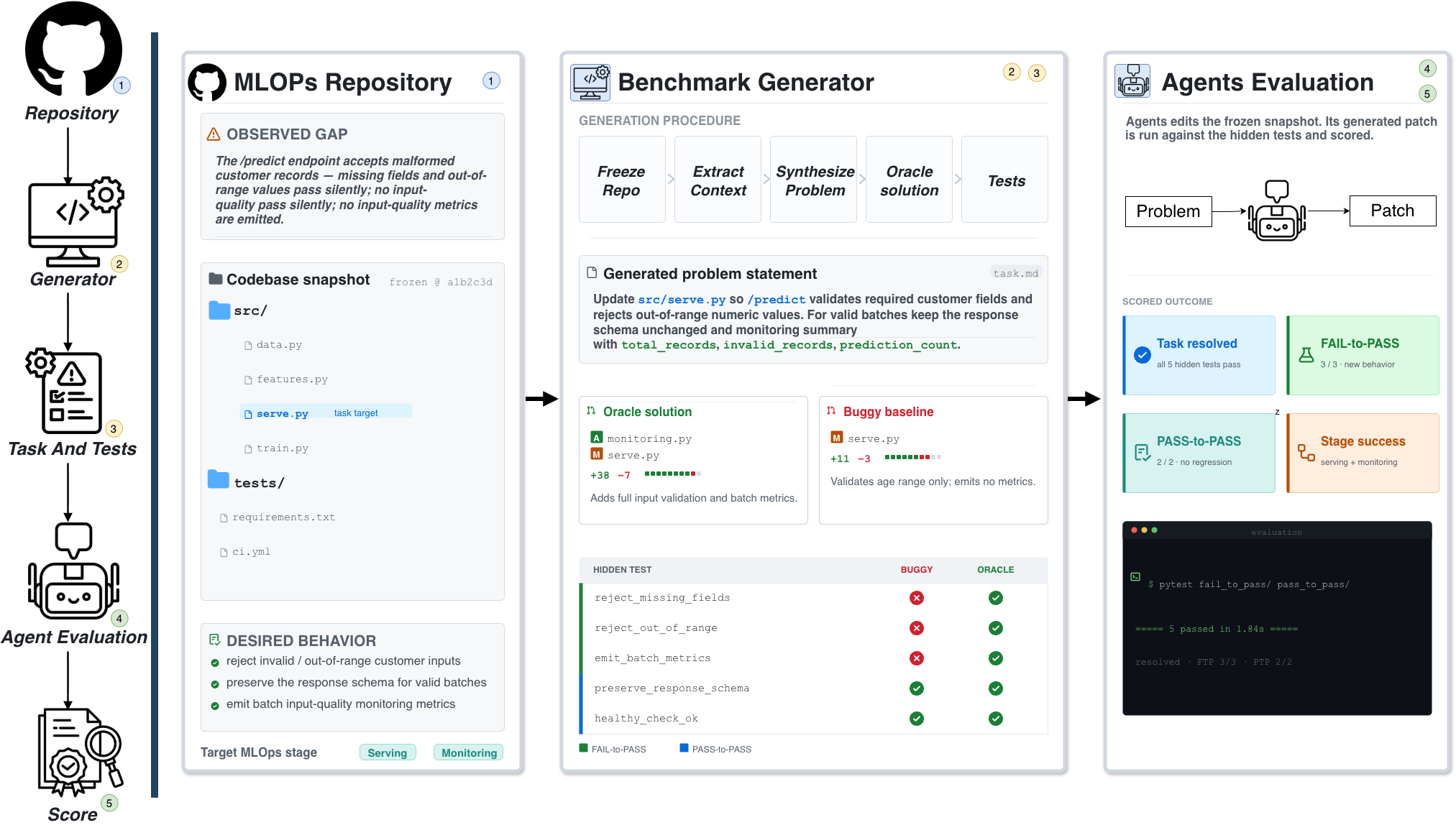}

\caption{\textsc{MLOps-Bench} construction and evaluation pipeline. Starting from a real ML repository, the generator freezes a source snapshot, extracts lifecycle evidence, synthesizes a repository-native MLOps task, constructs oracle and buggy patches, generates behavioral tests, and scores agents by applying their patches to the frozen repository.}
\label{fig:generation-diagram}
\end{figure*}


\section{Introduction}

Operating machine-learning systems in production is becoming a core expectation for AI agents, not just writing code or improving model accuracy. Production ML work differs from both ordinary software maintenance and competition-style ML: a correct change may need to preserve data contracts, maintain train/serve consistency, record artifact metadata, validate inference inputs, emit monitoring signals, or enforce governance gates. Existing benchmarks only partially measure these capabilities. SWE-bench evaluates repository-level issue resolution \citep{jimenez2024swebench}, while data-science and ML-agent benchmarks emphasize analysis, experimentation, or model improvement in task or competition settings \citep{jing2024dsbench,chan2024mlebench,huang2024mlagentbench, chen2025aiopslab, jha2025itbench}. They leave under-specified whether agents can make safe, executable MLOps changes inside existing ML repositories.

\textsc{MLOps-Bench} addresses this gap with a repository-grounded benchmark for production ML engineering. Its key novelty is that task validity is defined by lifecycle evidence rather than by issue availability, leaderboard performance, or generic programming difficulty: each task must be tied to native repository surfaces such as loaders, schemas, feature transforms, training configurations, model artifacts, serving paths, monitoring outputs, workflow files, or governance metadata. This grounding reflects known sources of ML technical debt and production-readiness risk \citep{kreuzberger2023mlops,sculley2015hidden,breck2017mltest}.

\textsc{MLOps-Bench} evaluates agents across six core MLOps stages plus one cross-stage integration category: data pipelines, feature engineering, model training, model serving, monitoring, CI/CD and governance, and end-to-end lifecycle integration. Each task pairs a repository-native request with executable behavioral tests, an oracle implementation, a plausible buggy baseline, and provenance metadata, testing whether agents satisfy production-relevant lifecycle contracts while preserving existing behavior and avoiding realistic failures: missing validation, train/serve skew, incomplete artifact metadata, malformed-input handling, ineffective monitoring thresholds, and non-enforcing governance checks. See Table~\ref{tab:ml-benchmark-comparison}.

\begin{table*}[t]
\centering
\small
\setlength{\tabcolsep}{4pt}
\renewcommand{\arraystretch}{1.2}

\newcommand{\rot}[1]{\rotatebox[origin=c]{70}{\textbf{\makecell{#1}}}}

\begin{tabular*}{\textwidth}{@{\extracolsep{\fill}}lccccccccc@{}}
\toprule
\textbf{Benchmark}
& \rot{Repo-grounded}
& \rot{Execution-based}
& \rot{Data}
& \rot{Feature Eng.}
& \rot{Training}
& \rot{Model Eval.}
& \rot{Model Serving}
& \rot{Monitoring}
& \rot{End-to-End}
\\
\midrule
DSBench \cite{jing2024dsbench}        & N & Y & Y & P & Y & Y & N & N & N \\
\addlinespace[2pt]
MLAgentBench  \cite{huang2024mlagentbench}  & Y & Y & Y & Y & Y & Y & N & N & P \\
\addlinespace[2pt]
MLE-bench \cite{chan2024mlebench}     & N & Y & Y & P & Y & Y & N & N & N \\
\addlinespace[2pt]
ML-Bench \cite{tang2023ml}      & Y & Y & P & P & P & Y & N & N & N \\
\addlinespace[2pt]
ML-Dev-Bench \cite{padigela2025ml}   & Y & Y & Y & P & Y & Y & N & P & P \\
\addlinespace[2pt]
\textsc{MLOps-Bench}     & Y & Y & Y & Y & Y & Y & Y & Y & Y \\
\bottomrule
\end{tabular*}

\caption{
Comparison of ML-agent and ML-engineering benchmarks.
Y = Yes, N = No, P = Partial.
}
\label{tab:ml-benchmark-comparison}
\end{table*}

We organize the benchmark around three research questions:\\
\textbf{RQ1: Repository-native MLOps capability.} Can AI agents implement production-oriented MLOps changes inside existing ML repositories while preserving native conventions and existing behavior?\\
\textbf{RQ2: Lifecycle-stage difficulty.} How does agent performance vary across data pipeline, feature engineering, model training, model serving, monitoring, CI/CD and governance, and end-to-end lifecycle tasks?\\
\textbf{RQ3: Behavioral robustness.} Do agents produce behaviorally correct MLOps implementations, or do they make superficially plausible changes that fail realistic production-oriented tests?

\noindent These RQs determine the benchmark design: RQ1 requires repository grounding, RQ2 requires lifecycle-stage coverage, and RQ3 requires executable tests that distinguish correct patches from plausible but incomplete MLOps changes. The contributions below follow directly from these requirements.

\noindent This paper makes the following contributions:

\textbf{1. A repository-grounded MLOps benchmark.}
We introduce \textsc{MLOps-Bench}, a benchmark generation pipeline that derives executable agent tasks from 31 public ML repositories \footnote{All selected repositories are released under open-source licenses.} spanning a combined 20 programming languages, frameworks, and tooling stacks, and 16 ML application domains.

\textbf{2. Lifecycle-stage evaluation for production ML engineering.}
We instantiate a taxonomy covering six core MLOps stages plus one cross-stage integration category: data pipelines, feature engineering, model training, model serving, monitoring, CI/CD and governance, and end-to-end lifecycle integration. The benchmark contains 212 validated tasks, enabling both aggregate and stage-level evaluation.

\textbf{3. Evidence-preserving task synthesis.}
We describe a context-to-requirement protocol that grounds each task in repository evidence, including native code paths, configuration files, tests, workflow files, model artifacts, monitoring utilities, and metadata. This design avoids generic MLOps prompts and ensures that each task is native to its source repository.

\textbf{4. Oracle, buggy-baseline, and behavioral-test construction.}
For each task, \textsc{MLOps-Bench} provides an oracle patch, a plausible buggy baseline, fail-to-pass tests, and pass-to-pass tests. In total, the benchmark includes 212 oracle patches, 212 buggy baselines, 636 fail-to-pass tests, and 856 pass-to-pass tests, designed to distinguish correct implementations from superficially plausible MLOps fixes.

\textbf{5. Auditable validation and human grounding checks.}
We attach provenance, licensing, snapshot, dependency, stage-label, requirement, oracle-rationale, and validation metadata to each instance. All released tasks pass validation checks, and a human audit comparing generated tasks with MLOps-candidate SWE-bench PRs supports constructing MLOps tasks directly from lifecycle evidence.

\textbf{6. Agent evaluation on production-oriented MLOps work.}
We evaluate 6 agents/models on \textsc{MLOps-Bench} and report task completion, test pass rate, stage-level performance, and categorized error analysis, showing how current coding and data-science agents perform on repository-native MLOps tasks.


\section{Related Work}

\paragraph{Repository-level software engineering benchmarks.}

Agent benchmarks have shifted from isolated code generation toward executable, repository-grounded evaluation. BugsInPy provides reproducible Python bugs for controlled debugging; SWE-bench tests whether agents resolve real GitHub issues by editing a codebase and passing tests \citep{10.1145/3368089.3417943,jimenez2024swebench}. This is the closest methodological precedent for MLOps-Bench: it treats the repository, not a standalone prompt, as the evaluation environment. Recent variants broaden this along language and modality axes: SWE-bench Multilingual, Multi-SWE-bench and SWE-bench Multimodal extend issue resolution across ecosystems and add visual artifacts \citep{swebenchmultilingual2025,zan2025multiswebench,yang2025swebenchmultimodal} yet their semantics remain centered on issue resolution. \textsc{MLOps-Bench} instead targets production ML lifecycle engineering, where tasks are validated using evidence such as data contracts, training artifacts, and monitoring signals.



\paragraph{Data-science and machine-learning agent benchmarks.}


A second line evaluates agents on data-science and ML workflows. DS-1000, DSEval, DSBench, DataSciBench, and DA-Code study code generation and increasingly realistic data-analysis tasks where agents inspect files, execute code, and recover from errors \citep{lai2023ds,zhang2024benchmarking,jing2024dsbench,zhang2025datascibench,huang-etal-2024-da}, but their outputs are analyses, notebooks, or modeling results rather than repaired operational ML systems.


Closer to ML engineering, MLAgentBench evaluates agents as iterative experimenters that write code, run experiments, inspect outputs, and revise their approach \citep{huang2024mlagentbench}; MLE-bench and MLE-Dojo scale this through Kaggle-derived or interactive ML-engineering tasks spanning data preparation, modeling, debugging, and repeated execution \citep{chan2025mle,qiang2026mle}; and ML-Bench and ML-Dev-Bench shift toward repository-level ML development \citep{tang2023ml,padigela2025ml}. MLOps-Bench complements these by shifting evaluation from model development or competition performance to production-oriented lifecycle changes inside existing repositories.

\paragraph{MLOps, ML reliability, and data-centric evaluation.}
MLOps research explains why production ML systems cannot be evaluated only as conventional software or isolated modeling problems. CRISP-ML(Q) frames ML development as a quality-assured process spanning data understanding, preparation, modeling, evaluation, deployment, and monitoring \citep{studer2021towards}, and MLOps surveys and architecture studies emphasize orchestration, versioning, CI/CD, continuous training, monitoring, and feedback loops \citep{kreuzberger2023mlops,9582569,10011505}. Such dependencies shape repository changes: a schema change can break feature computation, a feature fix can require retraining, retraining can invalidate evaluation thresholds, and a serving change can require updated monitoring, or governance checks.
Work on ML reliability and data-centric evaluation identifies many failure modes that motivate \textsc{MLOps-Bench}. Defects4ML collects reproducible ML faults, and studies of deep-learning bugs and ML testing show recurring problems around data assumptions, numerical instability, model construction, and library interactions \citep{morovati2023bugs,9000651,riccio2020testing}. DataPerf and ML data validation highlight failures in data selection, schema validation, and distribution monitoring, while pipeline-analysis tools show how preprocessing choices affect downstream model behavior \citep{MLSYS2019_928f1160,10.5555/3666122.3666357,10.1145/3589273}. These works motivate lifecycle-aware evaluation but typically study reliability as isolated bugs, data-quality tasks, or analysis settings rather than executable repository-native tasks for autonomous agents.


\paragraph{Positioning of \textsc{MLOps-Bench}.}
\textsc{MLOps-Bench} sits at the intersection of these lines: Repository-level software benchmarks provide the execution-based methodology, data-science and ML-agent benchmarks provide realistic development and experimentation settings, and MLOps and ML-reliability research identify the lifecycle dependencies and failure modes that make production ML difficult. No existing benchmark combines all three: repository-grounded executable evaluation, MLOps lifecycle semantics, and behavioral tests distinguishing correct fixes from plausible but incomplete production patches. \textsc{MLOps-Bench} fills this gap with tasks spanning data pipelines, feature engineering, model training, model serving, monitoring, CI/CD and governance, and end-to-end lifecycle integration.


\begin{table*}[!th]
\centering
\small
\begin{tabularx}{\textwidth}{p{0.15\textwidth}XXX}
\toprule
\textbf{Stage} & \textbf{Evidence} & \textbf{Generated requirement} & \textbf{Oracle surface} \\
\midrule
Data pipeline & Loaders, raw paths, parsers, schemas & Ingestion, validation, splitting, lineage, or quality behavior & Loaders, schema checks, preprocessing tests \\
Feature engineering & Feature functions, transforms, encoders, configs & Deterministic features, validation, or train/serve consistency & Feature modules, configs, unit tests \\
Model training & Training scripts, configs, metrics, checkpoints & Reproducibility, metric gates, checkpointing, or artifact metadata & Training loop, config parser, metrics tests \\
Model serving & API routes, batch inference, model loading, request schemas & Prediction behavior, input validation, artifact loading, or response contracts & Endpoint, inference wrapper, API tests \\
Monitoring & Logs, metrics, drift checks, dashboards, reports & Runtime metrics, drift checks, alert thresholds, or quality summaries & Monitoring utilities, emitters, regression tests \\
CI/CD and governance & Workflows, releases, Dockerfiles, model cards, metadata & Release gates, reproducibility metadata, policy checks, rollback records & Workflows, metadata, validation scripts \\
End-to-end & Evidence spanning at least two lifecycle stages & Artifact or decision flow across lifecycle boundaries & Native components plus integration tests \\
\bottomrule
\end{tabularx}
\caption{Generation schema for \textsc{MLOps-Bench} across lifecycle stages and end-to-end integration.}
\label{tab:stages}
\end{table*}

\section{Method}

\textsc{MLOps-Bench} generates executable benchmark tasks from real ML repositories through an evidence-preserving pipeline: repository selection, snapshotting, context extraction, stage-specific task generation, oracle construction, buggy-baseline construction, and executable validation. Each accepted task is traceable to repository evidence and evaluated by tests. Figure~\ref{fig:generation-diagram} summarizes the pipeline.

\subsection{Crawling Repositories}

We identify open-source ML repositories on GitHub using MLOps-related search queries, including \texttt{mlops}, \texttt{machine learning}, and \texttt{ml-ops}. Candidate repositories must be clonable, non-empty, and show evidence of ML engineering activity through source structure, issues or pull requests, workflow files, or pipeline artifacts. We then manually\footnote{These were done by ML engineers not involved in the inter-annotator activities.} select 31 repositories for lifecycle coverage, domain diversity, data and business context, buildability, and manageable audit size, excluding repositories with weak structure, limited ML pipeline signal, or poor fit for realistic MLOps-stage tasks. The resulting 212 tasks cover primarily Python-based ML systems, with additional coverage of YAML/config-heavy templates, TypeScript and web-serving code, shell automation, JavaScript utilities, and one C/C++ IoT repository. This scale is larger than SWE-Bench Lite's 11 repositories while remaining feasible for human audit.
 
\subsection{Repository Grounding and Context}
For each selected repository, we freeze a source snapshot and record provenance metadata, including URL, commit hash, license evidence, language, dependency manifests, test commands, workflow files, and likely modification surfaces. This supports reproducibility and auditability by preserving the source snapshot, dependency context, license evidence, and evaluation commands for each selected repository.

We then construct a context bundle with four views. The business context summarizes the domain, users, model purpose, and operational risk. The technical context identifies architecture, entrypoints, data paths, configuration files, model artifacts, service interfaces, and tests. The MLOps context records lifecycle evidence such as ingestion, feature computation, training, evaluation, serving, monitoring, CI, deployment, and governance. The requirement context converts observed gaps into candidate engineering requirements with relevant files, expected behavior, constraints, and preserved behavior.

This context prevents generic task generation. A monitoring task must be tied to an existing prediction path, metric output, logging surface, or deployment workflow; a data-validation task must be tied to an existing loader, schema assumption, or preprocessing step. Tasks without repository evidence, native implementation surfaces, or executable checks are rejected.

\subsection{Task Generation}

We generate tasks across six core MLOps stages plus one end-to-end integration category: data pipelines, feature engineering, model training, model serving, monitoring/observability, CI/CD and governance, and end-to-end MLOps flow. For each stage, the generator verifies repository evidence, derives a testable stage contract, identifies a minimal native modification surface, and writes a requirement that is specific enough to evaluate but does not reveal the implementation.

The end-to-end category is reserved for cross-boundary tasks rather than merely larger edits. Examples include training metadata consumed by serving, validation failures surfaced through CI, or monitoring summaries derived from batch inference outputs. Table~\ref{tab:stages} summarizes the stage evidence, generated requirements, and oracle surfaces.

\subsection{Problem Generation}

Each accepted problem is written as a repository-native engineering request with five fields. The business motivation explains why the change matters operationally. The technical context points to relevant components and current behavior. The requirement states the desired behavioral change in testable terms. The constraints specify repository conventions, allowed modification surfaces, compatibility requirements, and preserved behavior. The evaluation intent summarizes what hidden tests will check without revealing exact assertions. The generator stores the evidence used for each field so reviewers can audit grounding.

\subsection{Oracle, Buggy Baseline, and Tests}

After a problem is accepted, \textsc{MLOps-Bench} generates an oracle solution as one valid implementation of the benchmark contract. Oracle generation localizes the edit surface, applies the smallest native patch that satisfies the requirement, and records a rationale linking each changed file to the business motivation, technical context, and stage contract. The oracle is therefore a feasibility and validation artifact, not the only acceptable solution.

Following the controlled-defect tradition in software-testing benchmarks \citep{just2014defects4j}, the pipeline also generates a plausible buggy baseline for the same problem. The buggy patch should resemble a reasonable but incomplete engineering attempt: it may add the requested API but skip schema validation, compute a feature differently in training and serving, write a model artifact without metadata, handle normal inference but not malformed inputs, emit metrics without thresholds, or add a governance check that never fails. We avoid trivial baselines such as no-op patches or syntax errors because they do not test realistic agent failure modes.

Tests are generated after both patches exist. Fail-to-pass (F2P) tests encode the new benchmark contract and must fail on the buggy baseline but pass on the oracle. They target externally visible behavior such as rejected invalid records, stable feature values, saved artifact metadata, prediction response schemas, emitted monitoring summaries, or failing release gates. Pass-to-pass (P2P) tests protect behavior that should remain unchanged, usually by reusing or lightly extending existing tests around adjacent functionality. A benchmark is accepted only when the oracle passes both test sets, the buggy baseline fails at least one F2P test for the intended reason, and the tests do not depend on external services or hidden state.

\subsection{Validation and Evaluation}

A task is accepted only if grounding checks link the requirement to repository evidence, stage-contract checks confirm the claimed MLOps stage, behavioral tests distinguish the oracle from the buggy baseline, and P2P tests preserve existing behavior. During evaluation, an agent patch is applied to the frozen snapshot and scored using task completion rate, test pass rate, runtime, patch size, changed-file count, and error categories.

Benchmark construction used a separate development and evaluation split. We used two repositories and 25 tasks to iterate on generation prompts, oracle construction, test construction, and executable validation checks. The reported evaluation uses the remaining 29 repositories and 187 tasks.

\subsection{Dataset Statistics}

Overall, \textsc{MLOps-Bench} contains 212 tasks spanning six core MLOps stages and one end-to-end MLOps flow. Each task is grounded in a repository snapshot and scored with executable tests. The benchmark includes 25 development tasks and 187 evaluation tasks. In total, it contains 636 F2P tests, 856 P2P tests, and 1,492 tests overall. Table~\ref{tab:mlops-test-cases} shows the task and test distribution across stages. The per-stage counts are not exactly balanced because five candidate tasks were rejected during quality checks, primarily when the oracle did not pass reliably or the generated tests did not cleanly distinguish the oracle from the buggy baseline.

\begin{table}[t]
\centering
\small
\begin{tabular}{lrrrr}
\toprule
\textbf{MLOps Stage} & \textbf{Tasks} & \textbf{F2P} & \textbf{P2P} & \textbf{Total}  \\
\midrule
Data pipelines             & 31  & 93  & 120 & 213   \\
Feature engineering        & 30  & 90  & 118 & 208 \\
Model training             & 29  & 87  & 106 & 193  \\
Model serving              & 31  & 93  & 134 & 227 \\
Monitoring/observability   & 31  & 93  & 130 & 223   \\
CI/CD and governance           & 31  & 93  & 123 & 216   \\
End-to-end MLOps flow      & 29  & 87  & 125 & 212 \\
\midrule
\textbf{Dev}             & \textbf{25} & \textbf{73} & \textbf{120} & \textbf{193}  \\
\textbf{Evaluation}             & \textbf{187} & \textbf{563} & \textbf{736} & \textbf{1299}  \\
\textbf{Total}             & \textbf{212} & \textbf{636} & \textbf{856} & \textbf{1492}  \\
\bottomrule
\end{tabular}
\caption{Distribution of MLOps tasks and test cases across stages. F2P and P2P denote fail-to-pass and pass-to-pass tests, respectively.}
\label{tab:mlops-test-cases}
\end{table}

\subsection{Human Audit}
\subsubsection{Repository Issue Resue}
We audited whether existing repository-level issues can be repurposed as MLOps-stage tasks. Five annotators with ML engineering backgrounds double-annotated two sets of 25 items each: SWE-bench PRs flagged as MLOps candidates by our classification pipeline, and tasks produced by the \textsc{MLOps-Bench} generation pipeline. Each item was judged for whether it constitutes a valid MLOps task. 

Among the SWE-bench PRs, only 48\% (12/25) were unanimously confirmed as MLOps by both annotators, with 12\% rejected, 12\% unclear, and 28\% split. Annotator pooled agreement was moderate ($\kappa = 0.50$), indicating that the MLOps validity of repurposed PRs is contested even among domain-experienced reviewers. By contrast, annotators unanimously confirmed all 25 generated tasks as valid MLOps tasks.\footnote{We report raw agreement here because every generated task received the same label, which makes $\kappa$ degenerate for this binary judgment.} Although the sample is modest, this audit supports constructing MLOps tasks directly from explicit lifecycle evidence rather than repurposing generic repository issues.
\subsubsection{Test measure generalizable behavior}

To assess whether the F2P suites test the only specified behavior not generalizable behavior from oracle, we randomly selected 10 tasks and conducted a manual contract-coverage audit comparing each task’s stated behavioral contract with its corresponding F2P tests. Five annotators participated in the audit. Each task was independently evaluated by two annotators, yielding 20 task-level annotations in total. The annotators examined whether the F2P tests covered the important observable functional requirements in the task specification, including expected outputs, side effects, validation rules, edge cases, and integration behavior, rather than superficial or implementation-specific signals. The audit covered 4,633 lines of unique review material: 1,975 lines of F2P tests containing 40 test functions and 2,658 lines of structured task specifications. The annotators unanimously classified 9 of the 10 F2P suites as capturing the stated behavioral contract and 1 as not fully capturing it. Raw inter-annotator agreement was 100\%, with Cohen’s $\kappa$ = $1.0$. Even for the single suite that did not meet the binary coverage criterion, the two annotators’ independent estimates indicated that the tests covered an average of 85\% of the contract’s functional requirements. These results provide direct evidence that the F2P suites test the behaviors specified by the task contracts: 90\% fully met the independent contract-coverage criterion, and the remaining suite still covered a substantial majority of its stated functional requirements.

\subsubsection{Test accepting alternative correct solutions}
Five annotators participated in the study. Each task was independently reviewed by two annotators, yielding 20 task-level annotations in total, with each annotator evaluating four tasks. For every task, annotators examined the problem statement, F2P test suite, oracle patch, agent patch, and relevant repository context. The audit was substantial in terms of material reviewed. Because every task was double-annotated, this corresponds to an aggregate assigned review volume of 80,698 lines, in addition to any linked repository context inspected as needed. For 8 of the 10 patches, both annotators independently judged the agent patch to be a correct alternative implementation. The remaining two cases involved disagreement and were therefore reviewed by a third annotator, who classified both as correct alternatives. After adjudication, all patches were judged to satisfy the intended task contract while representing valid alternatives to the oracle.

\section{Results}

\noindent\textbf{A. Evaluation methodology:} To evaluate current models and coding systems, we group agents into two categories: coding agents and base agents, as shown in Table~\ref{tab:agent-types}.

\begin{table}[!th]
\centering
\small
\begin{tabularx}{\linewidth}{p{0.29\linewidth}X}
\toprule
\textbf{Agent type} & \textbf{Definition} \\
\midrule
Coding agent & An interactive coding system that can inspect files, edit the workspace, run commands, observe failures, and repair. Our coding agents are Codex (OpenAI \cite{openai2025codex}) and Claude Code (Anthropic \cite{anthropic2026claudecode}) \\
Base agent & A ReAct-style LangGraph agent with code generation and execution. \\
\bottomrule
\end{tabularx}
\caption{Agent types used in the evaluation.}
\label{tab:agent-types}
\end{table}

We instantiate each benchmark sample as a repository-level intervention rather than a prompt-only prediction problem. The evaluator materializes the baseline snapshot in an isolated workspace and constructs the task prompt from the benchmark specification. The evaluated agent accesses the workspace, edits files, and produces a patch. We then record the resulting patch, inject the scoring tests, and execute the test suite. To separate task solving from benchmark construction, evaluated agents receive only the task prompt and editable repository workspace; they do not receive repository-analysis metadata, oracle/buggy diffs, validation feedback, or generator retry traces. A run is considered successful only when the post-patch workspace satisfies all effective behavioral tests, while per-test outcomes, runtime, edit statistics, and failure categories are retained for further error analysis. This protocol compares systems by executable correctness under a common harness rather than by the plausibility of their natural-language plans.

\noindent\textbf{B. Performance Metrics: } We use task completion rate as the primary metric because a benchmark task is only solved when the agent produces a fully correct repository state that passes all scoring tests, while test pass rate captures partial progress and helps distinguish completely failed runs from solutions that implement some required behavior but miss edge cases or regressions.
\begin{table}[!th]
\centering
\small
\begin{tabularx}{\linewidth}{p{0.34\linewidth}X}
\toprule
\textbf{Metric} & \textbf{Definition} \\
\midrule
Task completion rate & Percentage of tasks for which all scoring tests pass. \\
Test pass rate & Percentage of individual scoring tests that pass across evaluated tasks. \\
\bottomrule
\end{tabularx}
\caption{Performance metrics.}
\label{tab:performance-metrics}
\end{table}
\begin{table*}[!th]
\centering
\small
\resizebox{\textwidth}{!}{%
\begin{tabular}{l l l r r}
\toprule
\textbf{Agent} & \textbf{Agent Type} & \textbf{Model} &
\textbf{Task Completion Rate (95\% CI)} &
\textbf{Test Pass Rate (95\% CI)} \\
\midrule
Codex & Coding agent & GPT-5.5
& 31.2\% [25.0, 43.1]
& 68.7\% [54.0, 75.6] \\

Claude Code (Sonnet) & Coding agent & Sonnet 4.6
& 16.5\% [8.6, 24.4]
& 52.3\% [44.9, 64.2] \\

Claude Code (Opus) & Coding agent & Opus 4.6
& 19.4\% [15.7, 25.8]
& 52.9\% [42.1, 63.9] \\

Base Agent Sonnet & Base agent & Claude Sonnet 4.6
& 14.5\% [7.8, 24.9]
& 39.2\% [25.9, 47.7] \\

Base Agent GPT 5.5 & Base agent & GPT-5.5
& 16.9\% [8.6, 25.9]
& 47.9\% [36.5, 52.7] \\

Base Agent GPT 5.2 & Base agent & GPT-5.2
& 13.4\% [8.8, 21.9]
& 31.4\% [24.7, 46.3] \\
\bottomrule
\end{tabular}%
}
\caption{Overall agent evaluation results.}
\label{tab:overall-results}
\end{table*}
\begin{figure*}[!ht]
\centering
\begin{minipage}[t]{0.48\textwidth}
  \centering
  \includegraphics[width=\linewidth]{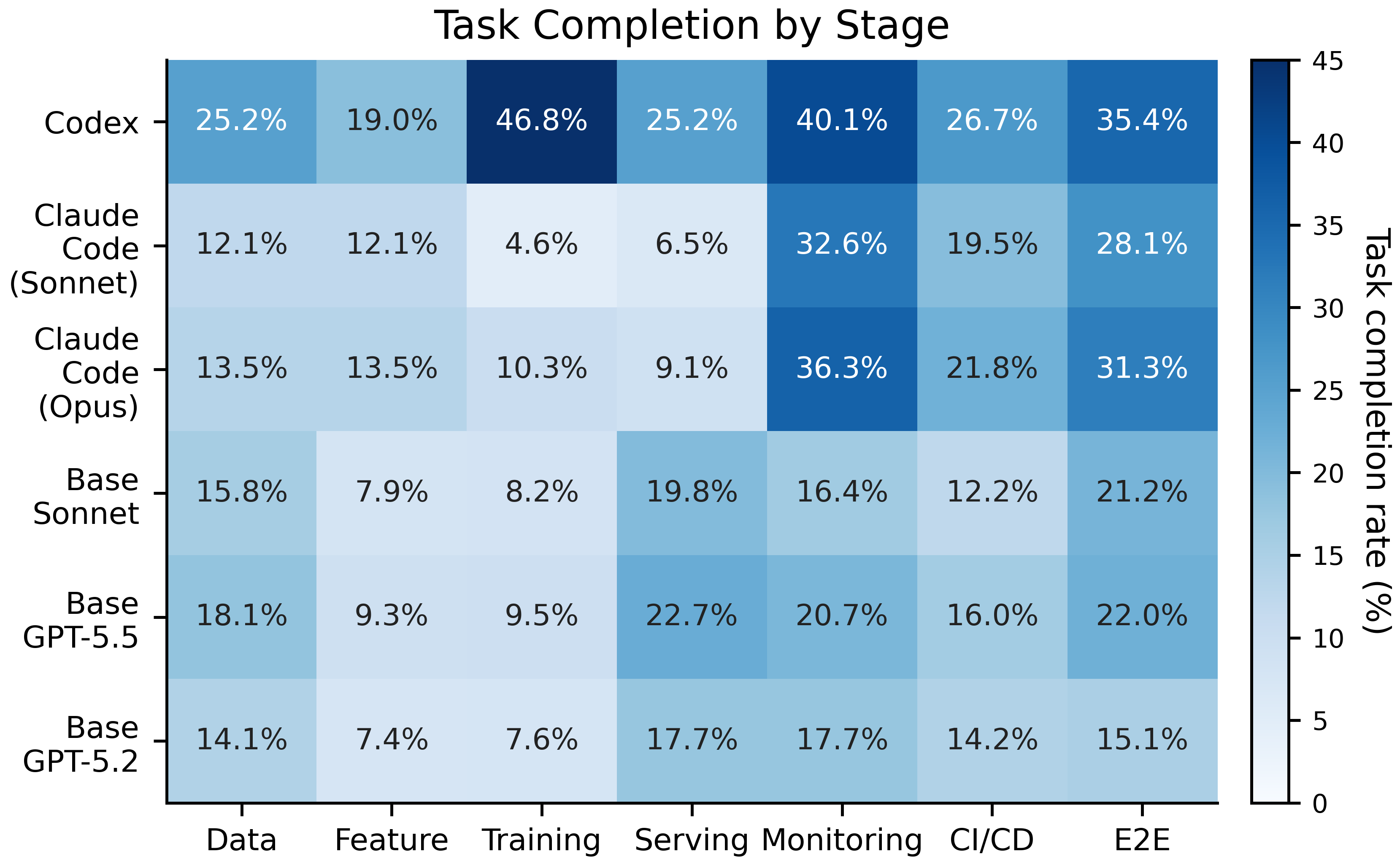}
  \small (a) Task completion rate by stage.
\end{minipage}\centering
\begin{minipage}[t]{0.48\textwidth}
  \centering
  \includegraphics[width=\linewidth]{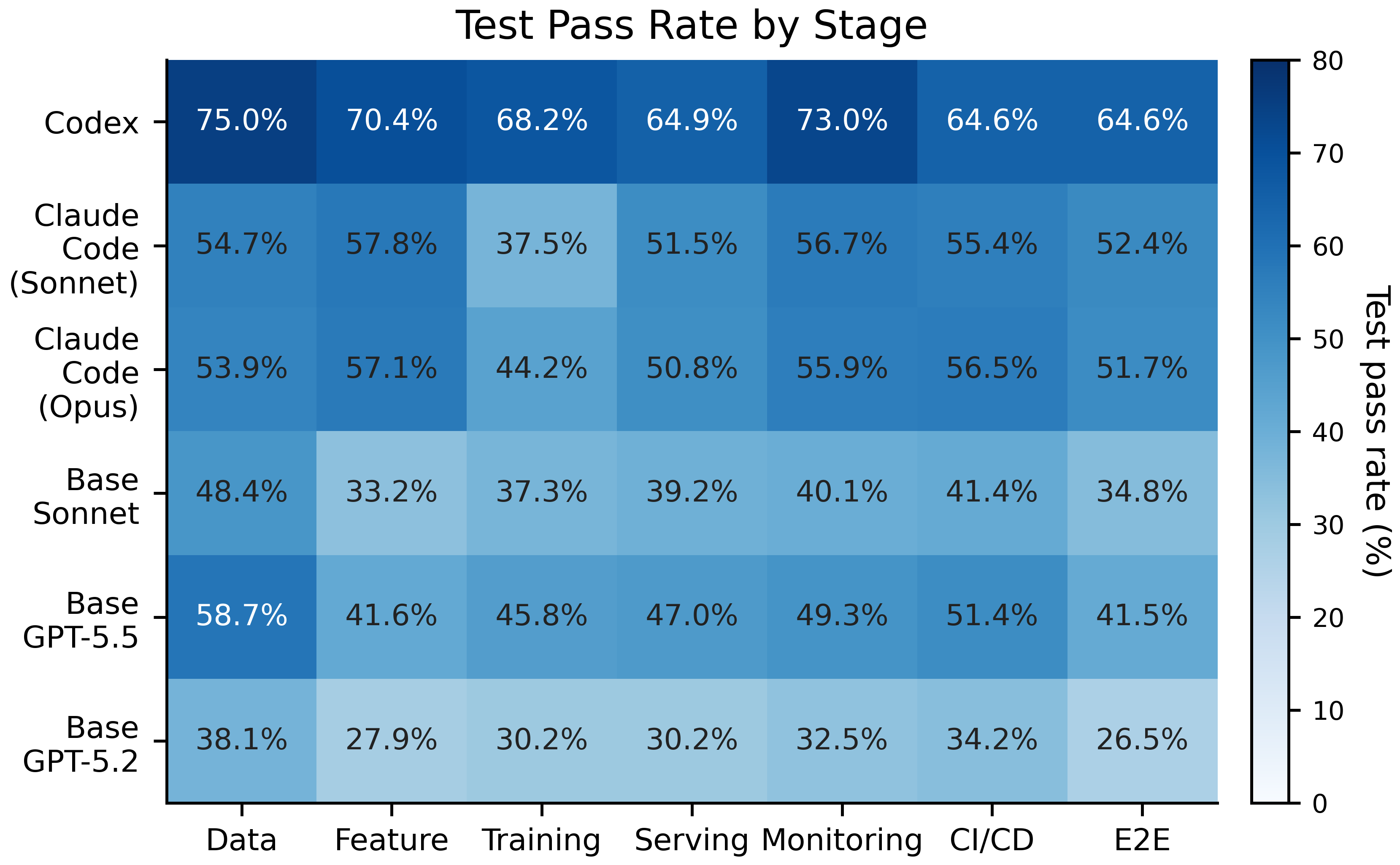}
  \small (b) Test pass rate by stage.
\end{minipage}

\caption{Stage-level performance heat maps across agents.}
\label{fig:stage-performance-heatmaps}
\end{figure*}
\begin{table*}[!th]
\centering
\small
\resizebox{\textwidth}{!}{%
\begin{tabular}{l r r r r r r}
\toprule
\textbf{Error Category} & \textbf{Codex} & \textbf{Claude Code (Sonnet)} & \textbf{Claude Code (Opus)} & \textbf{Base agent Sonnet} & \textbf{Base  Agent GPT 5.2} & \textbf{Base  Agent GPT 5.5} \\
\midrule
Wrong code functionalities & 56.8\% & 37.3\% & 42.0\% & 33.3\% & 60.4\% & 53.9\% \\
Fail/missing import & 20.5\% & 34.7\% & 20.0\% & 36.1\% & 21.3\% & 24.3 \%\\
Missing dependency & 13.6\% & 20.0\% & 14.0\% & 7.2\% & 15.6\% & 9.9 \%\\
Missing file/artifact & 4.5\% & 4.0\% & 10.0\% & 5.6\% & 6.2\% & 5.4 \% \\
Syntax & 3.4\% & 0.0\% & 0.0\% & 3.9\% & 4.7\% & 4.7\%\\
Config/data parse & 1.1\% & 1.3\% & 0.0\% & 0.0\% & 1.3\% &1.3\% \\
Test timeout & 0.0\% & 1.3\% & 14.0\% & 11.1\% & 0.5\%  & 0.5\%\\
\bottomrule
\end{tabular}%
}
\caption{Error distribution among failed tasks.}
\label{tab:error-analysis}
\end{table*}
We report the agent and model identifiers available at evaluation time. When a system exposes an agent product name rather than only a raw model name, we report both the product name and the underlying model identifier when available.

\noindent\textbf{C. Overall Results:} In this evaluation run, coding agents outperform the base-agent implementations on task completion. Codex obtains the highest task completion rate and test pass rate, while Claude Code (Opus) improves over Claude Code (Sonnet). The gap between task completion and test pass rate shows that many runs make substantial partial progress but fail at least one scoring test. 

Claude's use in benchmark generation does not imply that Claude models should be privileged during evaluation. Oracle and test generation are assisted synthesis steps: the generator has access to repository analysis, native-profile metadata, stage context, candidate problem specifications, oracle/buggy diffs, validation feedback, and retry loops. In contrast, the evaluated agent receives the task prompt and repository workspace, must infer the required implementation under tool and time constraints, and is scored only by the executable tests. Thus benchmark construction has substantially richer curated context than the agent solve setting.

\noindent\textbf{D. Results by MLOps Stage: } Each cell in Figure~\ref{fig:stage-performance-heatmaps} reports the task completion rate for that MLOps stage. Stage-level task completion is uneven. Monitoring/observability and end-to-end tasks show the strongest coding-agent performance, while feature engineering and serving remain difficult for several agents. The base agents trail the coding agents in this evaluation, reflecting the advantage of more comprehensive interactive coding workflows.

\noindent\textbf{E. Error Analysis:} Table~\ref{tab:error-analysis} reports the share of failed tasks assigned to each diagnostic category. Resolved tasks are excluded from this table.

Across agents, failures are dominated by partially correct but behaviorally incomplete implementations rather than simple inability to edit files. Codex most often produces executable patches that miss one or more task-specific requirements, with secondary failures from imports or newly introduced dependencies. Claude Code (Sonnet) fails more often at repository integration, especially broken imports, module paths, and dependency assumptions, while Claude Code (Opus) improves the success rate but has a broader spread of failures, including wrong functionality, missing artifacts, and timeouts from larger tool-loop edits. The one-shot base wrappers fail differently: without dynamic browsing, test execution, or repair, Base agent Sonnet frequently breaks package structure and imports, whereas Base Agent GPT 5.2 more often returns plausible but semantically insufficient patches that pass some tests but miss full task correctness.

\paragraph{Summary with respect to the research questions.}
For RQ1, current agents can solve a non-trivial fraction of repository-native MLOps tasks, but the best system completes only 31.2\% of evaluation tasks. For RQ2, performance varies across lifecycle stages, showing that aggregate scores hide stage-specific strengths and weaknesses. For RQ3, the gap between task completion and test pass rate, together with the error analysis, shows that many patches are partially correct but behaviorally incomplete.


\section{Conclusion}

We introduce \textsc{MLOps-Bench}, a repository-grounded benchmark for evaluating AI agents on production-oriented ML engineering tasks. Unlike Kaggle-style ML benchmarks or generic repository-repair benchmarks, \textsc{MLOps-Bench} targets lifecycle work that is central to operating ML systems, including data pipelines, feature engineering, model training, model serving, monitoring, CI/CD and governance, and end-to-end MLOps integration. Each task is grounded in real repository evidence and paired with executable tests, oracle patches, buggy baselines, and provenance metadata, enabling evaluation of both functional correctness and preservation of existing behavior.
Our evaluation suggests that current coding agents make meaningful progress on MLOps tasks but remain far from reliable production-level automation. In this run, the best-performing system achieves a 31.2\% task completion rate and a 68.7\% test pass rate, indicating that agents often implement partial or superficially plausible solutions while still missing important behavioral requirements. Stage-level and error analyses further confirm that failures are not limited to syntax or setup problems; many failures involve incorrect functionality, missing imports, dependency issues, or incomplete handling of repository-specific contracts. These results suggest that production ML engineering remains a challenging setting for AI agents and that future progress will require stronger repository understanding  and better lifecycle reasoning.

\section{Limitations}
\textsc{MLOps-Bench} is intended to evaluate production-oriented ML engineering in real repositories, but several limitations remain. First, \textsc{MLOps-Bench} is constructed from 31 public ML repositories selected for domain diversity, lifecycle evidence, and the ability to be reliably frozen, built, audited, and tested.  Although these repositories contain genuine ML lifecycle structure, they are open-source codebases rather than deployed commercial systems. As a result, \textsc{MLOps-Bench} measures production-style operational engineering in repository-native settings, but it should not be interpreted as fully representative of all industrial MLOps environments.

Second, the benchmark tasks are synthesized from repository evidence rather than collected exclusively from naturally occurring production incidents. This enables balanced lifecycle coverage, oracle solutions, buggy baselines, and executable fail-to-pass and pass-to-pass tests, but may introduce generation and test-design artifacts. Tasks can be more localized, explicit, and test-oriented than real MLOps requests, while finite test suites can only approximate production correctness: they cannot cover every downstream interaction, workload, data distribution, or operational failure mode.

Finally, because the generation pipeline relies on one model family while evaluation spans several, a potential concern is that the construction process could advantage same-family agents. We do not observe this in practice — the highest-scoring system (Codex, GPT-5.5) is not from the family used in generation — but we cannot fully rule out subtler stylistic alignment between generated tests and particular agents, and we flag it as a threat to validity that warrants further study.

\section{Ethical Considerations}

\textsc{MLOps-Bench} is constructed from public open-source repositories. We record provenance, source snapshots, dependency information, and license evidence for each selected project, and include only repositories whose licenses permit research and benchmark use under the corresponding license terms. Issues and pull requests are used only as public repository-level signals of activity and relevance; tasks are generated from lifecycle evidence in the source snapshot rather than from private user data or confidential production incidents.

We do not collect private information, credentials, secrets, or non-public organizational data, and we exclude repositories with obvious privacy, credential, or reproducibility concerns. The benchmark is intended for controlled evaluation of AI agents on frozen repositories, not for unsupervised modification of production ML systems. Since generated patches can be incomplete or unsafe outside the benchmark harness, benchmark success should not be interpreted as readiness for autonomous deployment without human review. We will release documentation covering construction, licensing, evaluation, and channels for reporting takedown, licensing, or data-quality concerns.

The authors used generative AI tools for limited language editing, rewriting suggestions, and assistance in improving exposition.  All scientific content, experimental design, code, analysis, claims, citations, and final text were reviewed, verified, and approved by the human authors.

\bibliography{custom}
\appendix

\section{Benchmark Example}
\label{sec:appendix-benchmark-example}

Table~\ref{tab:appendix-e2e-example} shows a representative MLOps-Bench instance from the end-to-end lifecycle category. The example illustrates how benchmark tasks require coordinated changes across data, feature, model, serving, monitoring, and release-governance surfaces rather than isolated code completion.

\begin{table*}[t]
\centering
\small
\begin{tabularx}{\textwidth}{p{0.20\textwidth}X}
\toprule
\textbf{Field} & \textbf{Example} \\
\midrule
Benchmark ID & \texttt{MLO-MLOPS\_AIRFLOW\_SAMPLE-07\_E2E\_FULL\_MLOPS\_FLOW-20260518\_113750} \\
Source repository & \texttt{mlops-airflow-sample} \\
MLOps stage & End-to-end full MLOps flow \\
Task title & Build an End-to-End MLOps Flow Across Data, Features, Training, Serving, Monitoring, and Governance \\
Existing stack & Airflow DAGs, Python task modules, MLflow metadata, KServe-style deployment YAML, Kubernetes manifests, and ELK observability manifests \\
\bottomrule
\end{tabularx}
\caption{Representative end-to-end MLOps-Bench task.}
\label{tab:appendix-e2e-example}
\end{table*}

Table~\ref{tab:appendix-e2e-example-json} formats the same benchmark content as structured JSON-style fields. This representation mirrors the metadata-rich task records used by MLOps-Bench while keeping the appendix human-readable.

\begin{table*}[t]
\centering
\small
\begin{tabularx}{\textwidth}{p{0.24\textwidth}X}
\toprule
\textbf{Field} & \textbf{JSON-style value} \\
\midrule
\texttt{business\_context} & \texttt{\{}\newline
\hspace*{1em}\texttt{"domain": "healthcare waiting-time forecasting",}\newline
\hspace*{1em}\texttt{"segments": ["specialty", "ROAZ region", "institution type", "urgency category"],}\newline
\hspace*{1em}\texttt{"users": ["operations teams", "capacity planners", "patient-routing coordinators"],}\newline
\hspace*{1em}\texttt{"risk": "bad releases can under-estimate high-demand waiting times, causing SLA misses, manual rework, and reputational damage",}\newline
\hspace*{1em}\texttt{"goal": "make each release auditable from raw data through deployed model and block unsafe releases before production traffic"}\newline
\texttt{\}} \\
\addlinespace
\texttt{task\_requirements} & \texttt{[}\newline
\hspace*{1em}\texttt{"create an Airflow-driven end-to-end release DAG",}\newline
\hspace*{1em}\texttt{"define a YAML release contract linking dataset version, feature version, MLflow run or model version, deployment manifest digest, monitoring manifest, rollback target, and approval status",}\newline
\hspace*{1em}\texttt{"add data-quality, feature-consistency, model-quality, serving-readiness, monitoring, and CI/CD governance gates",}\newline
\hspace*{1em}\texttt{"write a deterministic release evidence bundle with final decision blocked, canary\_only, or approved"}\newline
\texttt{]} \\
\addlinespace
\texttt{data\_quality\_constraints} & \texttt{\{}\newline
\hspace*{1em}\texttt{"dataset": "wachttijden",}\newline
\hspace*{1em}\texttt{"minimum\_rows": 10000,}\newline
\hspace*{1em}\texttt{"categorical\_feature\_completeness": ">99.5\%",}\newline
\hspace*{1em}\texttt{"target\_completeness": ">99.9\%",}\newline
\hspace*{1em}\texttt{"duplicate\_rate": "<0.5\%",}\newline
\hspace*{1em}\texttt{"unknown\_category\_rate": "<2\% per configured feature",}\newline
\hspace*{1em}\texttt{"feature\_freshness": "within 24 hours",}\newline
\hspace*{1em}\texttt{"model\_quality\_regression": "<=2\% versus baseline",}\newline
\hspace*{1em}\texttt{"p95\_serving\_latency": "<=500 ms",}\newline
\hspace*{1em}\texttt{"prediction\_error\_log\_rate": "<1\%"}\newline
\texttt{\}} \\
\bottomrule
\end{tabularx}
\caption{JSON-style representation of benchmark context, requirements, and constraints.}
\label{tab:appendix-e2e-example-json}
\end{table*}

This example shows why the suite is more than code completion: a successful solution has to coordinate data validation, feature engineering, model training, model serving, monitoring, rollback, and governance artifacts across the repository's native workflow.

\subsection{Example E2E MLOps Stage Flow}
\label{sec:appendix-e2e-flow}

The same benchmark encodes an end-to-end MLOps release as a sequence of artifact-producing and artifact-consuming stages. Table~\ref{tab:appendix-e2e-flow} summarizes the required handoffs.

\begin{table*}[t]
\centering
\small
\begin{tabularx}{\textwidth}{p{0.20\textwidth}X X}
\toprule
\textbf{Stage} & \textbf{Required Output} & \textbf{Consumed By} \\
\midrule
Data pipeline & Validated dataset version, data-quality report, schema/lineage metadata & Feature gate, model training gate, CI/CD evidence \\
Feature engineering & Feature version, feature statistics, training-serving skew baseline & Training, serving readiness, monitoring \\
Model training & MLflow run or model URI, evaluation metrics, promotion candidate status & Serving deployment gate, release governance \\
Model serving & Version-aware deployment manifest, readiness checks, prediction telemetry & Monitoring and rollback decisions \\
Monitoring/observability & Drift thresholds, latency/error thresholds, prediction-log contract & CI/CD governance gate \\
CI/CD governance & Release evidence bundle, final decision, rollback target & Production approval or blocked release \\
\bottomrule
\end{tabularx}
\caption{Artifact handoffs in the representative end-to-end MLOps benchmark.}
\label{tab:appendix-e2e-flow}
\end{table*}

For this task, the agent must implement the integration layer that verifies each handoff before deployment. A release can only be approved when the dataset, features, model artifact, serving manifest, monitoring contract, and rollback metadata all agree on the same release contract.

\section{Repository Counts by Profile and Workload}
\label{sec:appendix-repository-counts}

Programming languages and structured formats appear across the repositories as follows: Python in 29, JavaScript in 3, and TypeScript, C++, C/C++  in 1 repository each.

The ML workload table is a model-objective breakdown inferred from repository-analysis summaries and generated problem statements. Model-agnostic repositories are counted as platform/template/mixed rather than forced into a single classification or regression bucket.

\begin{table*}[t]
\centering
\small
\resizebox{\textwidth}{!}{%
\begin{tabular}{l r r r r p{0.42\textwidth}}
\toprule
\textbf{ML Workload Family} & \textbf{Source Repositories} & \textbf{Repository Share} & \textbf{Tasks} & \textbf{Task Share} & \textbf{Representative Scenarios} \\
\midrule
Classification / risk scoring & 17 & 54.8\% & 119 & 56.1\% & Pose/action, fruit, intent, sentiment, loan approval, churn, diabetes, predictive maintenance, Iris, employee attrition, regulated text \\
Regression / forecasting / pricing & 7 & 22.6\% & 49 & 23.1\% & Consignment pricing, diamond price, HDB resale price, energy demand, healthcare waiting time, order/demand forecasting \\
Platform/template/mixed MLOps & 4 & 12.9\% & 26 & 12.3\% & Kubernetes model serving, Azure ML solution templates, generic end-to-end MLOps templates, mixed ML portfolio tasks \\
Generative AI / LLM or image generation & 2 & 6.5\% & 11 & 5.2\% & MNIST CVAE image generation, wearable Ollama assistant \\
Recommendation / ranking & 1 & 3.2\% & 7 & 3.3\% & Amazon sales recommendation workflow \\
Total & 31 & 100.0\% & 212 & 100.0\% & \\
\bottomrule
\end{tabular}%
}
\caption{Repository and task counts by ML workload family.}
\label{tab:appendix-workload-counts}
\end{table*}

\subsection{Benchmark sample with Codex solutions}
\begin{figure*}[!ht]
\centering
\begin{minipage}[t]{0.40\textwidth}
  \centering
  \includegraphics[width=\linewidth]{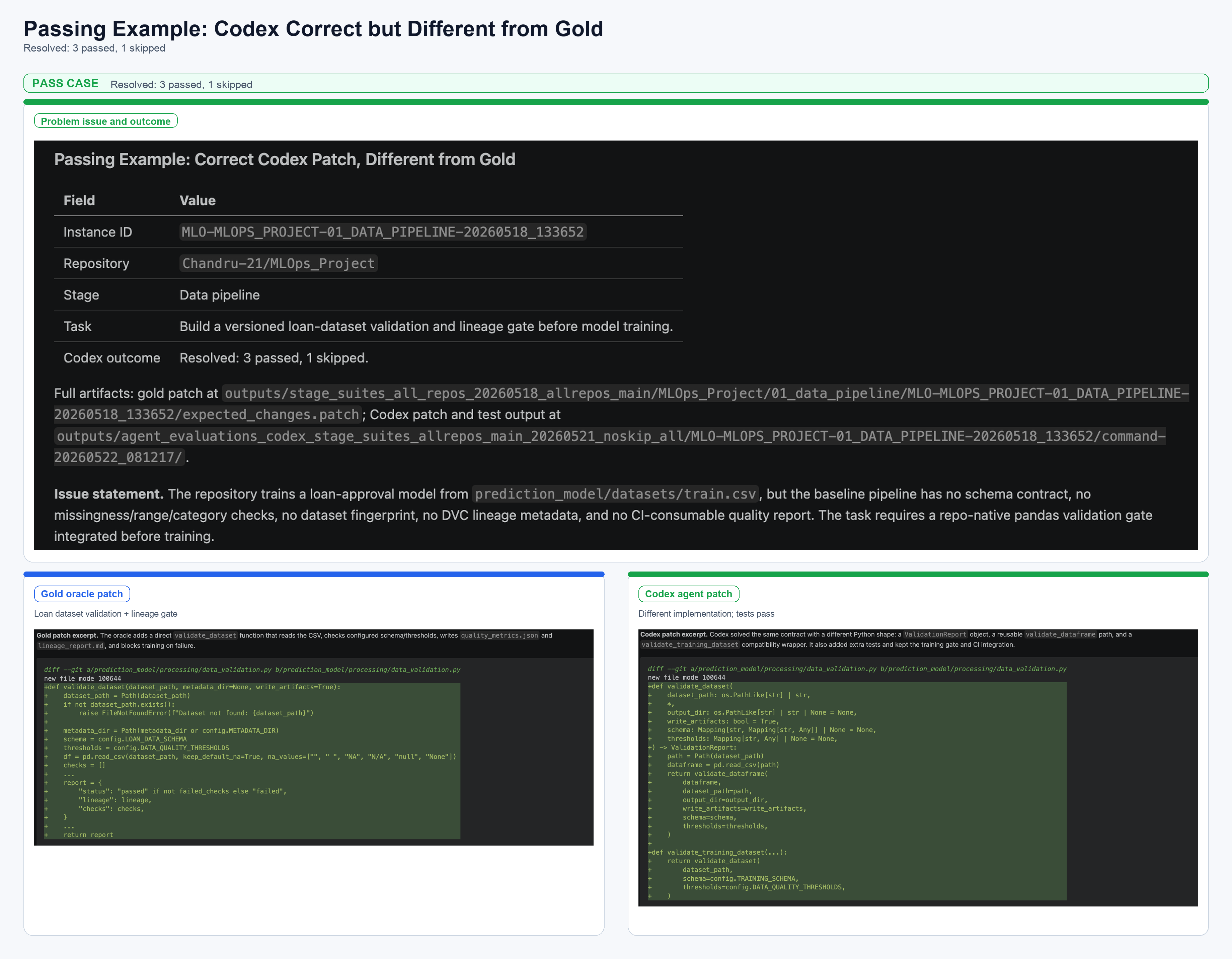}
  \small (a) Passing example.
\end{minipage}\centering
\begin{minipage}[t]{0.48\textwidth}
  \centering
  \includegraphics[width=\linewidth]{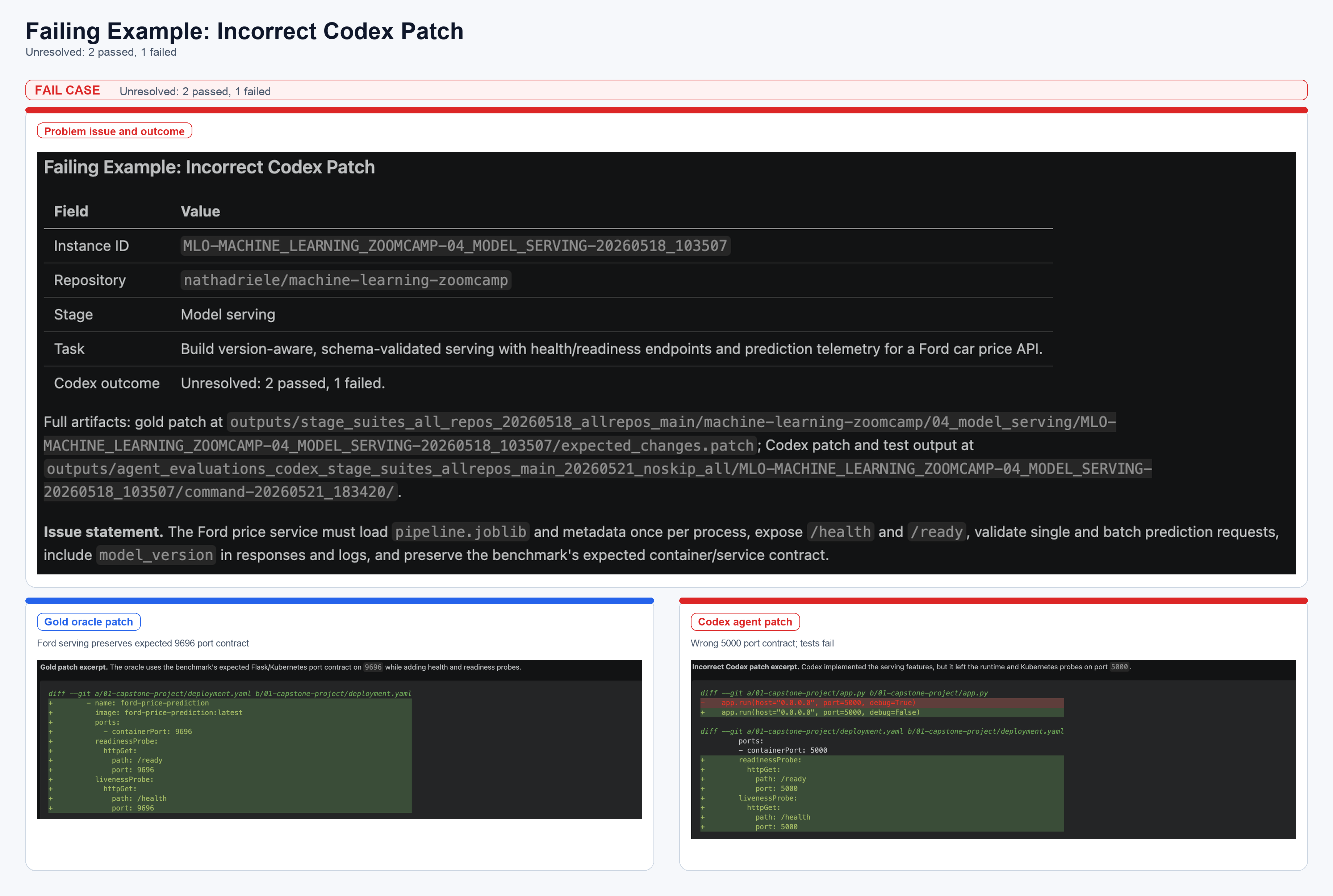}
  \small (b) Failing example.
\end{minipage}

\caption{Pass and fail scenarios from Codex on MLOps Bench}
\label{fig:pass-fail}
\end{figure*}

\section{Oracle Generation Prompt}
\label{sec:appendix-oracle-prompt}

Oracle solutions are generated with a repository-grounded prompt template. The full prompt includes the problem context, repository analysis, native repository profile, target stage contract, and source repository URL. A preview of the oracle instruction is shown in Figure~\ref{fig:appendix-oracle-prompt}.

\begin{figure*}[t]
\setlength{\FrameRule}{1pt} 
\begin{framed}
\begin{verbatim}
You are implementing the oracle solution for an MLOps benchmark task.

Inputs:
- Problem statement with business context, requirements, data context, and
  success criteria.
- Repository analysis, including native profile, existing files, dependency
  files, target MLOps stage contract, and preferred edit paths.
- Source repository URL for provenance.

Task:
Generate a problem-grounded repository patch that solves the task with working
repo-native code or configuration. Prefer extending existing files and patterns
over introducing generic templates.

Core constraints:
- Match the repository's native language, framework, build system, directory
  layout, and allowed file extensions.
- Reuse existing libraries, config systems, logging/telemetry patterns,
  entrypoints, and helper utilities.
- Do not add disallowed dependencies, unrelated services, or new deployment
  targets.
- Implement every problem requirement, not just a thematically related feature.
- For CI/CD and governance tasks, wire gates into a real workflow, script, or
  repository entrypoint.
- For end-to-end tasks, connect data pipeline, feature engineering, model
  training, model serving, monitoring, and governance through explicit artifact
  handoffs and release gates.
- Keep the patch focused: usually 2-6 repo-relative files and roughly 150-400
  lines total.

Output:
Return JSON containing:
- problem_grounding.requirement_coverage: one entry per original requirement,
  with exact requirement text, files changed, implemented behavior, and test
  obligation.
- problem_grounding.stage_contract_coverage: how the patch satisfies the target
  MLOps stage contract.
- files: complete repo-relative file contents for the oracle patch.
- integration_points: where the implementation is wired into the repository.
- repository_grounding: existing files/patterns extended.
- tests.test_cases: expected behavioral checks for fail-to-pass test generation.

\end{verbatim}
\end{framed}
\caption{Compact oracle-generation instruction used to produce repository-grounded reference solutions.}
\label{fig:appendix-oracle-prompt}
\end{figure*}

This prompt is intentionally stricter than a normal coding-agent prompt. The oracle generator receives repository-analysis metadata and must emit both code and an auditable grounding map; evaluated agents receive only the task prompt and editable repository snapshot.

\section{Human Audit Annotation Protocol}
\label{app:annotation}

This appendix documents the protocol used for the human audit of repository issue reuse. Annotators were given a shared rubric defining each MLOps lifecycle stage, together with accept/reject criteria. While the rubric supports a per-item stage judgment, the result reported in the main text aggregates to a single binary outcome: whether the problem statement constitutes a \emph{valid MLOps task}.

\subsection{What Annotators Reviewed}
Annotators reviewed two sets of 25 items each, drawn from different sources but judged under the same rubric. The first set consists of SWE-bench PRs that the automated pipeline flagged as MLOps candidates, each paired with the stage the pipeline assigned. For these, annotators judged whether the scraped problem statement genuinely describes MLOps lifecycle work, not whether the host repository is broadly an ML project. Annotators read the problem statement as scraped and consulted the linked issue or PR only to clarify problem context, not to infer implementation details absent from the statement itself, since GitHub often preserves formatting, code blocks, and screenshots that the spreadsheet view loses. The second set consists of problem statements produced by the \textsc{MLOps-Bench} generation pipeline, each likewise paired with its assigned stage. These were judged under the identical question—whether the statement constitutes a valid MLOps task at the assigned stage—so that generated and repurposed items are evaluated on the same criteria. To avoid bias, generated items were stripped of any pipeline metadata (oracle rationale, matched evidence, validation traces) and presented to annotators as standalone problem statements.

\subsection{MLOps stage Rubric}
The rubric below defines each stage with accept and reject criteria. It exists to make MLOps-validity judgments consistent across annotators; reviewers were instructed to apply their own data-science and MLOps judgment when a statement is ambiguous.

\paragraph{Data pipeline.} Ingestion, schema validation, data quality, dataset versioning, or lineage—data entering or moving through an ML workflow with explicit expectations about correctness, structure, quality, or traceability. \emph{Accept} when the issue concerns validating data shape/content, rejecting bad records, dataset versions, lineage, quality gates, or producing a clean dataset for downstream ML. \emph{Reject} when ``dataset'' only denotes an example used to reproduce a model/API bug, or for generic file parsing unrelated to data contracts.

\paragraph{Feature engineering.} Computing, reusing, materializing, or validating ML features reusable across training and inference. \emph{Accept} when the issue concerns feature definitions, transformations, offline/online consistency, freshness, or train/serve skew. \emph{Reject} ordinary preprocessing, tensor reshaping, or model-input conversion unless the statement explicitly concerns reusable features or train/serve consistency.

\paragraph{Model training.} Training pipelines, experiment tracking, model validation, and artifacts. \emph{Accept} when the issue concerns training entrypoints, trainer behavior, experiment logging, metric/artifact tracking, checkpointing, validation metrics, early stopping, or quality gates. \emph{Reject} when training is only part of a reproduction script for a generic framework bug; metrics appearing only in an error message are not experiment tracking.

\paragraph{Model serving.} Producing predictions in online, batch, or deployed contexts. \emph{Accept} when the issue concerns inference behavior, prediction endpoints, batch scoring, model loading for serving, latency, fallback behavior, health checks, prediction failures, or serving-time version selection. \emph{Reject} when \texttt{predict}/\texttt{inference} is used only as a library method in a local example.

\paragraph{Monitoring and observability.} Observing model or system behavior once data, training, or serving signals exist. \emph{Accept} when the issue concerns drift detection, model/service health, dashboards, alerts, production metrics, anomaly detection, or automated retraining triggers. \emph{Reject} when metrics are only training-time evaluation metrics, unit-test counts, or incidental logs.

\paragraph{CI/CD and governance.} Automation and controls deciding whether ML data, models, pipelines, or deployments are safe to release. \emph{Accept} when the issue concerns release automation, deployment gates, CI checks for ML artifacts/data/models, rollback behavior, audit trails, compliance metadata, or reproducibility gates. \emph{Reject} ordinary CI failures, dependency issues, or formatting checks unless they directly govern ML release safety.

\end{document}